\documentclass[12pt]{article}

\usepackage{axodraw}
\usepackage[totalwidth=480pt, totalheight=680pt]{geometry}

\makeatletter

\def\paragraph{\@startsection{paragraph}{4}{\z@}{+2.00ex plus
 +1ex minus +.2ex}{1.5ex plus .2ex}{\it\normalsize}}

\def\section{\@startsection {section}{1}{\z@}{+3.0ex plus +1ex minus
  +.2ex}{2.3ex plus .2ex}{\normalsize\bf\boldmath}}
\def\subsection{\@startsection{subsection}{2}{\z@}{+2.5ex plus +1ex
minus +.2ex}{1.5ex plus .2ex}{\normalsize\bf\boldmath}}
\def\subsubsection{\@startsection{subsubsection}{3}{\z@}{+3.25ex plus
 +1ex minus +.2ex}{1.5ex plus .2ex}{\normalsize\it}}

\expandafter\ifx\csname mathrm\endcsname\relax\def\mathrm#1{{\rm #1}}\fi

\newcounter{saveeqn}

\@addtoreset{equation}{section}

\newcount\@tempcntc
\def\@citex[#1]#2{\if@filesw\immediate\write\@auxout{\string\citation{#2}}\fi
  \@tempcnta\z@\@tempcntb\m@ne\def\@citea{}\@cite{\@for\@citeb:=#2\do
    {\@ifundefined
       {b@\@citeb}{\@citeo\@tempcntb\m@ne\@citea
        \def\@citea{,\penalty\@m\ }{\bf ?}\@warning
       {Citation `\@citeb' on page \thepage \space undefined}}%
    {\setbox\z@\hbox{\global\@tempcntc0\csname
b@\@citeb\endcsname\relax}%
     \ifnum\@tempcntc=\z@ \@citeo\@tempcntb\m@ne
       \@citea\def\@citea{,\penalty\@m}
       \hbox{\csname b@\@citeb\endcsname}%
     \else
      \advance\@tempcntb\@ne
      \ifnum\@tempcntb=\@tempcntc
      \else\advance\@tempcntb\m@ne\@citeo
      \@tempcnta\@tempcntc\@tempcntb\@tempcntc\fi\fi}}\@citeo}{#1}}

\def\@citeo{\ifnum\@tempcnta>\@tempcntb\else\@citea
  \def\@citea{,\penalty\@m}%
  \ifnum\@tempcnta=\@tempcntb\the\@tempcnta\else
   {\advance\@tempcnta\@ne\ifnum\@tempcnta=\@tempcntb \else
\def\@citea{--}\fi
    \advance\@tempcnta\m@ne\the\@tempcnta\@citea\the\@tempcntb}\fi\fi}

\def\asymp#1%
{\mathrel{\raisebox{-.4em}{$\widetilde{\scriptstyle #1}$}}}

\def\Nequal#1%
{\mathrel{\raisebox{-.5em}{$\stackrel{=}{\scriptstyle\rm#1}$}}}
\newcommand{\dsl}[1]{\not \hspace{-0.7mm}#1}
\def\dsl{\mathpalette\make@slash}
\def\make@slash#1#2{\setbox\z@\hbox{$#1#2$}%
  \hbox to 0pt{\hss$#1/$\hss\kern-\wd0}\box0}

\def\beq{\begin{equation}}
\def\eeq{\end{equation}}
\def\beqar{\begin{eqnarray}}
\def\eeqar{\end{eqnarray}}
\def\barr#1{\begin{array}{#1}}
\def\earr{\end{array}}
\def\bfi{\begin{figure}}
\def\efi{\end{figure}}
\def\btab{\begin{table}}
\def\etab{\end{table}}
\def\bce{\begin{center}}
\def\ece{\end{center}}
\def\nn{\nonumber}
\def\disp{\displaystyle}
\def\text{\textstyle}

\def\eps{\epsilon}

\def\refeq#1{\mbox{(\ref{#1})}}

\def\reffi#1{\mbox{Figure~\ref{#1}}}

\def\refse#1{\mbox{Section~\ref{#1}}}

\def\refapp#1{\mbox{App.~\ref{#1}}}
\def\citere#1{\mbox{Ref.~\cite{#1}}}
\def\citeres#1{\mbox{Refs.~\cite{#1}}}

\newcommand{\ri}{{\mathrm{i}}}
\newcommand{\rd}{{\mathrm{d}}}

\def\mathswitchr#1{\relax\ifmmode{\mathrm{#1}}\else$\mathrm{#1}$\fi}

\newcommand{\PW}{\mathswitchr W}

\newcommand{\PZ}{\mathswitchr Z}

\newcommand{\PH}{\mathswitchr H}

\newcommand{\Pe}{\mathswitchr e}

\newcommand{\Pt}{\mathswitchr t}

\newcommand{\Pep}{\mathswitchr {e^+}}
\newcommand{\Pem}{\mathswitchr {e^-}}

\def\mathswitch#1{\relax\ifmmode#1\else$#1$\fi}

\newcommand{\MW}{\mathswitch {M_\PW}}

\newcommand{\MZ}{\mathswitch {M_\PZ}}
\newcommand{\MH}{\mathswitch {M_\PH}}
\newcommand{\Me}{\mathswitch {m_\Pe}}

\newcommand{\Mt}{\mathswitch {m_\Pt}}

\newcommand{\sing}{{\mathrm{sing}}}

\newcommand{\soft}{{\mathrm{soft}}}
\newcommand{\coll}{{\mathrm{coll}}}

\def\Li{\mathop{\mathrm{Li}_2}\nolimits}

\def\draftdate{\relax}
\def\mda{\relax}
\def\mua{\relax}
\def\mla{\relax}
\def\draft{
\def\thtystars{******************************}
\def\sixtystars{\thtystars\thtystars}
\typeout{}
\typeout{\sixtystars**}
\typeout{* Draft mode!
         For final version remove \protect\draft\space in source file *}
\typeout{\sixtystars**}
\typeout{}
\def\draftdate{\today}
\def\mua{\marginpar[\boldmath\hfil$\uparrow$]%
                   {\boldmath$\uparrow$\hfil}%
                    \typeout{marginpar: $\uparrow$}\ignorespaces}
\def\mda{\marginpar[\boldmath\hfil$\downarrow$]%
                   {\boldmath$\downarrow$\hfil}%
                    \typeout{marginpar: $\downarrow$}\ignorespaces}
\def\mla{\marginpar[\boldmath\hfil$\rightarrow$]%
                   {\boldmath$\leftarrow $\hfil}%
                    \typeout{marginpar: $\leftrightarrow$}\ignorespaces}
\def\Mua{\marginpar[\boldmath\hfil$\Uparrow$]%
                   {\boldmath$\Uparrow$\hfil}%
                    \typeout{marginpar: $\uparrow$}\ignorespaces}
\def\Mda{\marginpar[\boldmath\hfil$\Downarrow$]%
                   {\boldmath$\Downarrow$\hfil}%
                    \typeout{marginpar: $\downarrow$}\ignorespaces}
\def\Mla{\marginpar[\boldmath\hfil$\Rightarrow$]%
                   {\boldmath$\Leftarrow $\hfil}%
                    \typeout{marginpar: $\leftrightarrow$}\ignorespaces}
\overfullrule 5pt
\oddsidemargin -15mm
\marginparwidth 29mm
}

\def\stars{\strut\leaders\hbox{*}\hfill\strut}
\def\starline{\hfil\strut\hfil\hbox to \textwidth {\stars}\hfil}

\begin{document}
\thispagestyle{empty}
\def\thefootnote{\fnsymbol{footnote}}
\setcounter{footnote}{1}
\null
\draftdate
\strut\hfill MPP-2003-61\\
\strut\hfill hep-ph/0308246 
\vfill
\begin{center}
{\large \bf\boldmath
Separation of soft and collinear singularities \\[.5em]
from one-loop $N$-point integrals
\par} \vskip 2em
\vspace{1cm}
{\large
{\sc Stefan Dittmaier} } 
\\[.5cm]
{\it  Max-Planck-Institut f\"ur Physik (Werner-Heisenberg-Institut) \\
             F\"ohringer Ring 6, D-80805 M\"unchen, Germany}
\par 
\end{center}\par
\vfill
\vskip 2.0cm {\bf Abstract:} \par 
The soft and collinear singularities of general scalar and tensor
one-loop $N$-point integrals are worked out explicitly. As a result
a simple explicit formula is given that expresses the singular part
in terms of 3-point integrals. Apart from predicting the singularities, 
this result can be used to transfer singular one-loop integrals from one
regularization scheme to another or to subtract soft and collinear 
singularities from one-loop Feynman diagrams directly in momentum space.
\par
\vskip 1cm
\noindent
August 2003   
\null
\setcounter{page}{0}
\clearpage
\def\thefootnote{\arabic{footnote}}
\setcounter{footnote}{0}

\section{Introduction}
\label{se:intro}

Many interesting high-energy processes at future colliders, such as
the LHC and an $\Pep\Pem$ linear collider, lead to final states with
more than two particles, rendering precise predictions much more
complicated than for $2\to2$ particle reactions. 
The necessary calculation of radiative corrections bears additional 
complications and requires a further development of calculational
techniques, as recently reviewed in \citere{Dittmaier:2003sc}.

Full one-loop calculations for processes with more than two
final-state particles require, for instance, the
evaluation of scalar and tensor $N$-point integrals. 
For $N=5,6$ several approaches have been proposed in the literature
\cite{Me65,vanNeerven:1983vr,vanOldenborgh:1989wn,Denner:2002ii}.
In this context the two main complications concern a numerically
stable evaluation of tensor integrals on the one hand and 
a proper separation of infrared (soft and collinear) singularities on
the other.
In \citere{Denner:2002ii} the direct reduction of scalar 5-point to
4-point integrals, as proposed in \citere{Me65}, has been extended to
tensor integrals. The nice feature in this approach is the avoidance
of leading inverse Gram determinants, which necessarily appear in the
well-known Passarino--Veltman reduction \cite{Passarino:1979jh}
and lead to numerical instabilities at the phase-space boundary.

In this paper we focus on the treatment of infrared (IR)
or so-called ``mass singularities'' at one loop.
According to Kinoshita \cite{Kinoshita:ur}, such mass singularities
arise from two configurations that both lead to logarithmic
singularities. {\it Collinear} singularities appear if a massless
external particle splits into two massless internal particles of a loop
diagram, and {\it soft} singularities arise if two external (on-shell)
particles exchange a massless particle. If the involved particles
are not precisely massless, the corresponding singularities show
up as large logarithms involving the small masses, like $\ln(\Me/Q)$ where
$\Me$ is the electron mass and $Q$ a large scale. 
If the masses involved in the singular configurations are exactly zero,
the singularities appear as
regularized divergences, such as $1/\eps$ poles where
$\eps=(4-D)/2$ in $D$-dimensional regularization. In both cases
an analytical control over such terms is highly desirable, either
in order to perform cancellations at the analytical level or
to carry out resummations. In the following we show how to
extract mass singularities from general tensor $N$-point integrals
and finally give an explicit result for the singular parts in terms
of related 3-point integrals.
Moreover, we describe several ways how this result can be 
exploited and give some examples illustrating the easy use of
our final result.
The method to derive the general result of this paper was already 
applied in \citere{Beenakker:2002nc} to specific 5-point integrals,
which appear in the next-to-leading order QCD corrections to the
processes $gg/q\bar q\to\Pt\bar\Pt\PH$.

Mass singularities of the virtual and real radiative corrections
are intrinsically connected in field theory. In fact, the famous
Kinoshita--Lee--Nauenberg (KLN) theorem \cite{Kinoshita:ur,Lee:1964is}
states that the singularities completely cancel in sufficiently
inclusive quantities. The result of this paper allows for a simple
analytical handling of the mass singularities of the virtual one-loop
corrections. The singular structure of the real corrections induced
by one-parton emission (which corresponds to the one-loop level)
can be easily read from so-called subtraction formalisms
\cite{Ellis:1980wv} which are designed for separating these singularities.
Using these results and the KLN theorem, the one-loop singular
structure of complete QCD and SUSY-QCD amplitudes has been derived
in a closed form in \citere{Catani:2001ef}.
Note that the attitude of this paper is rather different, since the
one-loop integrals are inspected themselves, eventually leading to
a prescription for extracting the singularities diagram by diagram.

The paper is organized as follows:
In \refse{se:classification} we set our conventions and describe the
situations in which mass singularities appear at one loop.
Section~\ref{se:separation} contains the actual separation of the mass
singularities and our final result at the end of the section.
In \refse{se:disc+appl} we describe various ways to make use of the result
and present some explicit applications.
Section~\ref{se:summary} contains a short summary.
In the appendices we present a proof of an auxiliary identity used in
\refse{se:separation} and provide a list of mass-singular scalar
3-point integrals that frequently appear in applications.

\section{\boldmath{One-loop $N$-point integrals and mass singularities}}
\label{se:classification}

We consider the general one-loop $N$-point integrals
\beq 
T^{(N)}_{\mu_1\dots\mu_P}(p_0,\dots,p_{N-1},m_0,\dots,m_{N-1}) =
\frac{(2\pi\mu)^{(4-D)}}{\ri\pi^{2}}\int\!\rd^{D}q\,
\frac{q_{\mu_1}\cdots q_{\mu_P}}{N_0\cdots N_{N-1}},
\label{eq:Npointint}
\eeq
with the denominator factors
\beq \label{D0Di}
N_n= (q+p_n)^{2}-m_n^2+\ri 0, \qquad n=0,\ldots,N-1.
\eeq
A diagrammatic illustration is shown in \reffi{fig:TN}.
\bfi
\centerline{
{ \unitlength 1pt
\begin{picture}(200,125)(10,-15)
\Line(20, 85)( 60, 85)
\Line(20,  5)( 60,  5)
\Line(60, 85)(140, 85)
\Line(60,  5)(140,  5)
\Line(60, 85)( 60,  5)
\Vertex( 60, 85){3}
\Vertex( 60,  5){3}
\Vertex(140,  5){3}
\Vertex(140, 85){3}
\Line(140, 85)(180, 85)
\Line(180,  5)(140,  5)
\DashLine(160, 25)(140, 25){3}
\DashLine(160, 65)(140, 65){3}
\DashLine(140,  5)(140, 85){3}
\LongArrow(50,35)(50,55)
\LongArrow(90,94)(110,94)
\LongArrow(110,-2)(90,-2)
\put(10,43){$q+p_n$}
\put( 85,-17){$q+p_{n-1}$}
\put( 85,105){$q+p_{n+1}$}
\put(64,43){$m_n$}
\put(95,11){$m_{n-1}$}
\put(95,73){$m_{n+1}$}
\put(170,40){$\vdots$}
\end{picture}
}
}
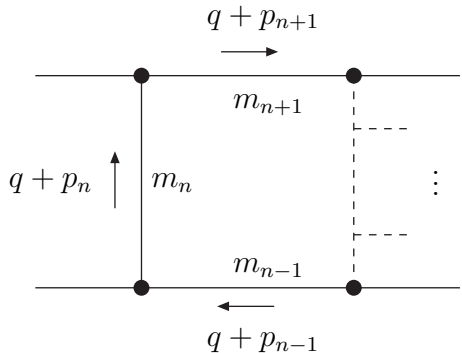
\caption{Momentum and mass assignment for a general one-loop $N$-point 
integral \refeq{eq:Npointint}}
\label{fig:TN}
\efi
Note that we do not set the momentum $p_0$ to zero, as it is often done
by convention, but keep this variable in order to facilitate a generic
treatment of related integrals. 
In particular, with this convention all $T^{(N)}_{\dots}$ are 
invariant under exchange of any two propagator denominators $N_n$, or
equivalently of two pairs of $(p_n,m_n)$.
We follow the usual convention to denote $N$-point integrals with
$N=1,2,\dots$ as
\beq
T^{(1)}\equiv A, \quad
T^{(2)}\equiv B, \quad
T^{(3)}\equiv C, \quad
T^{(4)}\equiv D, \quad
T^{(5)}\equiv E, \dots .
\eeq
Whenever the index $n$ on momenta $p_n$ or masses $m_n$ exceeds the range
$n=0,\ldots,N-1$, it is understood as modulo $N$, i.e.\ $n=0$ and $n=N$
are equivalent, etc.
For later use, we introduce the variables
\beq
c_{mn} = 2(p_{n+1}-p_n)(p_m-p_n), \qquad
d_{mn} = (p_m-p_n)^2-m_m^2.
\eeq

We are interested in ``mass'' singularities that appear if combinations
of external squared momenta, $(p_{n+1}-p_n)^2$, and internal masses $m_n$
become small, but do not consider singular configurations that are related
to specific or isolated points in phase space, such as thresholds
or forward scattering.
We can, thus, distinguish two sets of parameters: one set that comprises
all quantities $(p_m-p_n)^2$, $m_n$ with fixed non-zero values, and
another set of those quantities that are considered to be small,
i.e.\ which formally tend to zero.
In order to simplify the notation, we define an operation, indicated
by a caret~$\hat{}$ over a quantity $X$, which implies that all
small quantities are set to zero in $\hat X$.

As shown by Kinoshita \cite{Kinoshita:ur}, ``mass'' (or ``IR'') singularities
can appear in one-loop diagrams in the following two situations:
\begin{enumerate}
\item
An external line with a light-like momentum (e.g.\ a massless external 
on-shell particle) is attached to two massless propagators, i.e.\
there is an $n$ with
\beq
(p_{n+1}-p_n)^2\to0,  \quad m_{n+1}\to0, \quad m_n\to0 \qquad
\Rightarrow \quad \hat d_{n,n+1}=\hat d_{n+1,n}=0.
\label{eq:collcond}
\eeq
The singularity is logarithmic and originates from integration momenta
$q$ with
\beq
q\to-p_n+x_n(p_n-p_{n+1}),
\label{eq:qcolllimit}
\eeq
where $x_n$ is an arbitrary real variable. Since the momentum
$(q+p_n)$ on line $n$ is then collinear to the external momentum
$(p_n-p_{n+1})$, such singularities are called {\it collinear} singularities.
\item
A massless particle is exchanged between two on-shell particles, i.e.\
there is an $n$ with
\beqar
&& m_n\to0, \quad (p_{n-1}-p_n)^2-m_{n-1}^2\to0, 
\quad (p_{n+1}-p_n)^2-m_{n+1}^2\to0
\nn\\
&& \qquad \Rightarrow \quad \hat d_{n-1,n}=\hat d_{n+1,n}=0.
\label{eq:softcond}
\eeqar
The singularity is also logarithmic and originates from integration momenta
$q$ with
\beq
q\to-p_n,
\label{eq:qsoftlimit}
\eeq
i.e.\ the momentum transfer of the $n$th propagator tends to zero.
Therefore, these singularities are called {\it soft} singularities.
\end{enumerate}

In the following we focus on integrals with $N>3$ and express the
singular structure of one-loop $N$-point integrals in terms
of 3-point integrals which are easily calculated with standard
techniques, as for instance described in 
\citeres{vanOldenborgh:1989wn,Passarino:1979jh,'tHooft:1978xw}.
Of course, the same is true for the cases $N=1,2$, i.e.\ for 
tadpole and self-energy integrals, which are even simpler.

\section{\boldmath{Separation of mass singularities}}
\label{se:separation}

In this section, we first consider the asymptotic behaviour of the
denominator of the integrand in Eq.~\refeq{eq:Npointint} in the individual
collinear and soft limits. Based on these partial results we derive
a simple expression that resembles the whole integrand in all
singular regions. Applying the loop integration to this expression 
directly leads to our main result which expresses the singular
structure of an arbitrary $N$-point integral \refeq{eq:Npointint} with
$N>3$ in terms of 3-point integrals.

\subsection{Asymptotic behaviour in collinear regions}

For an integration momentum $q$ in the collinear domain,
as specified in Eq.~\refeq{eq:qcolllimit},
the two propagator denominators $N_n$ and $N_{n+1}$ tend to zero
($N_n,N_{n+1}\to0$), and the $N_k$ behave as
\beq
N_k \sim \left[-p_n+x_n(p_n-p_{n+1})+p_k\right]^2-m_k^2
=N_n + m_n^2 - x_n c_{kn} + d_{kn} 
\quad \mbox{for} \quad k\ne n,n+1.
\hspace{2em}
\eeq
The collinear limit is mass singular if the external momentum squared
$(p_{n+1}-p_n)^2$ and the two masses $m_n$, $m_{n+1}$ are small.
In this limit the two propagator denominators $N_n$, $N_{n+1}$ tend to
zero, but the others remain finite (for $x_n\ne0,1$):
\beq
N_k \to - x_n \hat c_{kn} + \hat d_{kn}
\qquad \mbox{for} \quad k\ne n,n+1.
\eeq
Note that the variable $x_n$ is the only integration variable that is
not fixed by the collinear limit. The product of all regular
propagators $N_k^{-1}$ can be decomposed into a sum over these
propagators via taking the partial fraction,
\beq
\prod_{k=0 \atop k\ne n,n+1}^{N-1} \frac{1}{N_k} \sim
\prod_{k=0 \atop k\ne n,n+1}^{N-1} \frac{1}{- x_n \hat c_{kn} + \hat d_{kn}} 
= \sum_{k=0 \atop k\ne n,n+1}^{N-1} \frac{A^\coll_{nk}}{- x_n \hat c_{kn} + \hat d_{kn}}
= \sum_{k=0 \atop k\ne n,n+1}^{N-1} \frac{A^\coll_{nk}}{N_k}.
\label{eq:partialfrac}
\eeq
The coefficients $A^\coll_{nk}$ are functions of the variables
$\hat d_{kn}$ and $\hat c_{kn}$ alone and thus fixed by the external kinematics.
The explicit result for $A^\coll_{nk}$ reads
\beq
A^\coll_{nk} = \frac{\hat c_{kn}^{N-3}}{\disp\prod_{l=0 \atop l\ne k,n,n+1}^{N-1}
(\hat c_{kn}\hat d_{ln}-\hat c_{ln}\hat d_{kn})},
\label{eq:Ank}
\eeq
as proven in \refapp{app:partialfrac}.
The collinear singularity arises from the region where the propagator
denominators $N_n$ and $N_{n+1}$ both become small with no preference
to any of the two. In order to reveal this equivalence, we rewrite
$A^\coll_{nk}$ using 
\beqar
c_{kn} &=& 2(p_{n+1}-p_n)(p_k-p_n)
\nn\\
&=& \left[(p_k-p_n)^2-m_k^2\right]
-\left[(p_k-p_{n+1})^2-m_k^2\right]
+(p_{n+1}-p_n)^2
\nn\\
\Rightarrow\quad \hat c_{kn}
&=& \hat d_{kn} - \hat d_{k,n+1} \qquad \mbox{for} \quad (p_{n+1}-p_n)^2\to0
\eeqar
and 
\beq
\hat c_{kn}\hat d_{ln}-\hat c_{ln}\hat d_{kn} 
= \hat d_{kn} \hat d_{l,n+1} - \hat d_{ln} \hat d_{k,n+1}.
\eeq
Inserting these relations, we get
\beq
A^\coll_{nk} = \frac{(\hat d_{kn} - \hat d_{k,n+1})^{N-3}}%
{\disp\prod_{l=0 \atop l\ne k,n,n+1}^{N-1}
(\hat d_{kn} \hat d_{l,n+1} - \hat d_{ln} \hat d_{k,n+1})},
\eeq
in which the equivalence of the $n$th and $(n+1)$th propagators
is evident.

Multiplying Eq.~\refeq{eq:partialfrac} with $N_n^{-1}N_{n+1}^{-1}$
yields a relation, which is valid in the collinear limit,
between the product of all $N$ propagators and a linear combination
of products $N_n^{-1}N_{n+1}^{-1}N_k^{-1}$ involving only three
propagators,
\beq
\prod_{k=0}^{N-1} \frac{1}{N_k} \;\sim\;
\sum_{k=0 \atop k\ne n,n+1}^{N-1}
\frac{A^\coll_{nk}}{N_n N_{n+1} N_k}.
\label{eq:collasymp}
\eeq
Thus, the collinear singularity associated with the
propagators $N_n^{-1}$, $N_{n+1}^{-1}$ in an $N$-point
integral is expressed in terms of a sum of 3-point integrals
involving the $n$th, the $(n+1)$th, and any other line of the diagram.

\subsection{Asymptotic behaviour in soft regions}

The soft singularity connected with a massless propagator $N_n$ arises
from momenta $q\to -p_n$, where $N_{n-1}, N_n, N_{n+1}\to0$.
The other denominators tend to a regular limit in this case,
\beq
N_k \to (p_k-p_n)^2-m_k^2 = \hat d_{kn}
\qquad \mbox{for} \quad k\ne n-1,n,n+1,
\eeq
and the product of all propagators behaves like
\beq
\prod_{k=0}^{N-1} \frac{1}{N_k} \;\sim\;
\frac{A^\soft_{n}}{N_{n-1} N_n N_{n+1}}
\qquad \mbox{with} \quad 
A^\soft_n = \prod_{l=0 \atop l\ne n-1,n,n+1}^{N-1} \frac{1}{\hat d_{ln}}.
\label{eq:softasymp}
\eeq
We still have to consider the possibility that one or both ends of
the soft line $n$ is part of a collinear configuration treated above.
If this is the case, the soft limit can be reached as limiting case
of a collinear limit. Assuming $n$ again as the soft line, the two
``degenerate'' collinear limits are $x_n\to0$ and $x_{n-1}\to1$.
Both lead to $q\to-p_n$, but in the former case lines $n$ and $n+1$
correspond to a collinear configuration, in the latter lines
$n-1$ and $n$.
It is quite easy to see that the soft asymptotic behaviour
\refeq{eq:softasymp} is already correctly included in the collinear
behaviour \refeq{eq:collasymp} in either case, because
\beq
A^\coll_{n,n-1}=A^\coll_{n-1,n+1} = A^\soft_n
\qquad \mbox{for} \quad \hat d_{n-1,n}=\hat d_{n+1,n}=0.
\eeq

\subsection{Final result}

From the above considerations it is clear that we obtain an expression for
the asymptotic behaviour of the product of all propagator denominators
in all collinear and soft regions
upon adding the asymptotic expressions of all collinear and soft regions,
which can be read from Eqs.~\refeq{eq:collasymp} and \refeq{eq:softasymp},
and carefully avoiding double-counting of soft asymptotic terms.
To this end, we define
\beq
A_{nk} = \left\{ \barr{ll} 
A^\coll_{nk} &\mbox{if Eq.~\refeq{eq:collcond} is fulfilled, but}\\
             &\mbox{neither $[k=n-1$ and $\hat d_{n-1,n}=0]$}\\
             &\mbox{nor $[k=n+2$ and $\hat d_{n+2,n+1}=0]$,}\\[.5em]
A_n^\soft    &\mbox{if $k=n-1$ and Eq.~\refeq{eq:softcond} is fulfilled,}\\[.5em]
0\qquad      &\mbox{otherwise.}
\earr \right.
\eeq
With this definition we can write down the asymptotic behaviour
valid for all collinear and soft regions as
\beq
\prod_{k=0}^{N-1} \frac{1}{N_k} \;\sim\;
\sum_{n=0}^{N-1}
\sum_{k=0 \atop k\ne n,n+1}^{N-1}
\frac{A_{nk}}{N_n N_{n+1} N_k}.
\label{eq:asymp}
\eeq
Obviously each soft part is included by the $A_n^\soft$ terms exactly once,
and the collinear contributions from
$A^\coll_{n,n-1}$ and $A^\coll_{n-1,n+1}$ are omitted if they are already
covered by the soft terms $A_n^\soft$.
Integrating Eq.~\refeq{eq:asymp} over 
$\frac{(2\pi\mu)^{(4-D)}}{\ri\pi^2}\int\rd^D q$ on the l.h.s.\
yields the scalar integral $T_0^{(N)}$ and on the r.h.s.\
a linear combination of scalar 3-point integrals $C_0$, which has 
exactly the same structure of collinear and soft singularities as
$T_0^{(N)}$.
An analogous relation is obtained for tensor integrals if the 
additional factor $q_{\mu_1}\cdots q_{\mu_P}$ is included in the integration,
since this factor does not lead to additional singularities. 
Note that the asymptotic relation \refeq{eq:asymp}, which 
describes the leading behaviour, is in fact sufficient
to extract all mass singularities from the one-loop integral, since
the degree of the singularities is logarithmic.
In summary the complete mass-singular part of a general
one-loop tensor $N$-point function reads
\beqar 
\lefteqn{ \hspace*{-2em}
T^{(N)}_{\mu_1\dots\mu_P}(p_0,\dots,p_{N-1},m_0,\dots,m_{N-1})\Big|_\sing
}
\nn\\[.3em]
&=&
\sum_{n=0}^{N-1}
\sum_{k=0 \atop k\ne n,n+1}^{N-1}
A_{nk} \, C_{\mu_1\dots\mu_P}(p_n,p_{n+1},p_k,m_n,m_{n+1},m_k).
\label{eq:Npointsing}
\eeqar
The sum over $n$ and $k$
runs over all subdiagrams whose scalar integral develops a
collinear or soft singularity. A tensor integral is, however, not
necessarily mass singular if the related scalar integral develops such
a singularity. For such tensor integrals the regular 3-point integrals
on the r.h.s.\ of Eq.~\refeq{eq:Npointsing} could be dropped.
We note that for $P\ge2$, artificial ultraviolet singularities appear
in the tensor 3-point integrals on the r.h.s.\ of Eq.~\refeq{eq:Npointsing}.
These can be regularized in dimensional regularization and easily
separated from the mass singularities (see, e.g., the appendix of
\citere{Denner:2002ii}). 

In order to render the above result more useful, we present a list of
mass-singular $C_0$ functions in \refapp{app:C0s}. This paper, thus,
contains the needed ingredients to predict the mass singularities
of most scalar $N$-point functions occurring in practice. 
To obtain the singularities
of tensor integrals, only the 3-point tensor integrals have to be derived,
which can be easily inferred with the well-known Passarino--Veltman
algorithm \cite{Passarino:1979jh} (see also 
\citeres{vanOldenborgh:1989wn,Denner:2002ii}).

\section{\boldmath{Discussion and applications}}
\label{se:disc+appl}

\subsection{Possible applications of the final result}

The relation \refeq{eq:Npointsing} can be exploited in various directions:
\begin{itemize}
\item
As pointed out in the previous section, the mass singularities of
arbitrary $N$-point integrals can be easily derived from 3-point functions.
This statement is true in any regularization scheme, i.e.\ for any
$N$-point integral all small-mass logarithms and/or poles in
$(D-4)$ in dimensional regularization can be easily inferred.
\item
The singular integral $T^{(N)}_{\mu_1\dots\mu_P}|_\sing$ can be used to
translate any IR-divergent $N$-point integral from one regularization
scheme to another. To this end, the regularization-scheme-independent
difference 
\beq
T^{(N)}_{\mu_1\dots\mu_P} - T^{(N)}_{\mu_1\dots\mu_P}|_\sing
\;\;=\;\; \mbox{(IR finite)}
\label{eq:TNdiff}
\eeq
is considered. For the translation from one scheme to the other
only the singular part $T^{(N)}_{\mu_1\dots\mu_P}|_\sing$,
and thus the relevant 3-point integrals, have to be known in the two
regularization schemes. 
\item
The trick described in the last item has been used in 
\citere{Beenakker:2002nc} to translate $D$-dimensional 5-point integrals
into a mass regularization with $D=4$, in order to make use of the
direct reduction \cite{Me65,Denner:2002ii} 
of 5-point to 4-point integrals, which works in four space-time
dimensions. In this context it was observed that the formal relation 
between 5-point and 4-point integrals, which was derived in four
dimensions, is also valid in $D$ dimensions up to ${\cal O}(D-4)$ terms, 
since the extraction
of the singularities works in any regularization scheme with the same
linear combination of 3-point integrals. 
From the results of this paper we conclude that this statement generalizes to
arbitrary $N$-point integrals, i.e.\ the reduction of an $N$-point integral
to 4-point integrals works in $D$ dimensions in precisely the same way
as in four dimensions [up to terms of ${\cal O}(D-4)$], 
without the appearance of extra terms.
\item
Since Eq.~\refeq{eq:Npointsing} has been derived in momentum space,
it could also be used as local counterterm in the momentum-space
integral, i.e.\ taking the difference in Eq.~\refeq{eq:TNdiff} before the
integration over the loop momentum $q$, the integral becomes
IR (soft and collinear) finite and can be evaluated without IR regulator.
The loop integration of the subtracted part is extremely simple, because
it involves only 3-point functions, and can be added again after
the integration of the difference.
This procedure could be very useful in purely numerical approaches
to loop integrals, as e.g.\ described in \citere{Ferroglia:2002mz}.

In other words, Eq.~\refeq{eq:Npointsing} can serve as the basis of
a subtraction formalism for one-loop corrections, very similar to
the frequently used subtraction formalisms for real corrections
as worked out in \citere{Ellis:1980wv}.
\end{itemize}

\subsection{Sudakov limit of the one-loop box integral}

As a simple application, we consider the box integral
$D_{\dots}(p_0,p_1,p_2,p_3,m_0,m_1,m_2,m_3)$ in the so-called Sudakov limit,
where all external squared momenta and internal masses are considered
to be much smaller than the two Mandelstam variables
\beq
s=(p_2-p_0)^2, \quad t=(p_3-p_1)^2.
\eeq
In this limit, there are four regions for soft singularities, and 
Eq.~\refeq{eq:Npointsing} yields
\beqar
\lefteqn{ \hspace{-2em}
D_{\dots}(p_0,p_1,p_2,p_3,m_0,m_1,m_2,m_3)\Big|_\sing }
\nn\\
&=& \phantom{{}+{}}
\frac{1}{s} \, C_{\dots}(p_1,p_2,p_3,m_1,m_2,m_3)
+\frac{1}{t} \, C_{\dots}(p_2,p_3,p_0,m_2,m_3,m_0)
\nn\\ && {}
+\frac{1}{s} \, C_{\dots}(p_3,p_0,p_1,m_3,m_0,m_1)
+\frac{1}{t} \, C_{\dots}(p_0,p_1,p_2,m_0,m_1,m_2).
\label{eq:Dsudakov}
\eeqar
For the scalar integral
this is in agreement with Eq.~(57) of \citere{Roth:1996pd},
where the remaining finite contribution was derived as well.
In \citere{Roth:1996pd} also tensor integrals up to rank 4 have been
considered; the singularities predicted by Eq.~\refeq{eq:Dsudakov}
have been checked against these results.
The soft singularities in the $C_{\dots}$ functions on the r.h.s.\
of Eq.~\refeq{eq:Dsudakov} arise from integration momenta
$q\to p_2$, $p_3$, $p_0$, $p_1$, respectively.
The singular terms in the tensor coefficients of the $C_{\dots}$ functions
can be related to the respective scalar $C_0$ functions rather easily.
For instance, shifting the integration momentum $q\to q-p_2$ in
the first $C_{\dots}$ function on the r.h.s.\
of Eq.~\refeq{eq:Dsudakov}, power-counting in the shifted 
momentum $q$ shows that terms with $q$ in the numerator are not mass singular.
Thus, in the first tensor $C_{\dots}$ function 
only covariants built of the momentum
$p_2$ alone receive singular coefficients that are 
all proportional to the respective $C_0$ function.
The same reasoning applies to the other three $C_{\dots}$ functions.
In summary, the mass-singular terms of the tensor 4-point functions in the
Sudakov limit are given by
\beqar
\lefteqn{ 
D^{\mu_1\dots\mu_P}(p_0,p_1,p_2,p_3,m_0,m_1,m_2,m_3)\Big|_\sing }
\nn\\ 
&\sim& \phantom{{}+{}}
 (-1)^P\, \frac{p_2^{\mu_1}\cdots p_2^{\mu_P}}{s}\, C_0(p_1,p_2,p_3,m_1,m_2,m_3)
+(-1)^P\, \frac{p_3^{\mu_1}\cdots p_3^{\mu_P}}{t}\, C_0(p_2,p_3,p_0,m_2,m_3,m_0)
\nn\\ && {}
+(-1)^P\, \frac{p_0^{\mu_1}\cdots p_0^{\mu_P}}{s}\, C_0(p_3,p_0,p_1,m_3,m_0,m_1)
+(-1)^P\, \frac{p_1^{\mu_1}\cdots p_1^{\mu_P}}{t}\, C_0(p_0,p_1,p_2,m_0,m_1,m_2),
\nn\\
\label{eq:Dsudakov2}
\eeqar
where the $\sim$ sign indicates that regular terms 
have been dropped, i.e.\ Eq.~\refeq{eq:Npointsing} is not applied 
literally.

\subsection{Singular structure of some 5-point integrals}

\paragraph{{A 5-point integral for the process $gg\to\Pt\bar\Pt\PH$}}

In \citere{Beenakker:2002nc} the three different types of IR-singular
5-point integrals that appear in the next-to-leading order (NLO) QCD 
correction to $gg/q\bar q\to\Pt\bar\Pt\PH$ have been calculated
in dimensional regularization.
One of the corresponding pentagon diagrams is shown on the
l.h.s.\ of \reffi{fig:ggtthdiag}.
\bfi
\centerline{
{ \unitlength 1pt
\begin{picture}(200,125)(10,-15)
\Gluon(20, 85)( 60, 85){4}{4}
\Gluon(20,  5)( 60,  5){4}{4}
\Gluon(60, 85)(140, 85){4}{8}
\Gluon(60,  5)(140,  5){4}{8}
\Gluon(60, 85)( 60,  5){4}{8}
\Vertex( 60, 85){3}
\Vertex( 60,  5){3}
\Vertex(140,  5){3}
\Vertex(140, 45){3}
\Vertex(140, 85){3}
\DashLine(140,45)(180,45){5}
\ArrowLine(180, 85)(140, 85)
\ArrowLine(140, 85)(140, 45)
\ArrowLine(140, 45)(140,  5)
\ArrowLine(140,  5)(180,  5)
\LongArrow(47,55)(47,35)
\LongArrow(149,55)(149,75)
\LongArrow(149,15)(149,35)
\LongArrow(110,97)(90,97)
\LongArrow(90,-5)(110,-5)
\put(36,43){$q$}
\put( 85,-17){$q+p_1$}
\put(155,22){$q+p_1-p_3$}
\put(155,62){$q+p_4-p_2$}
\put( 85,105){$q-p_2$}
\put(69,43){$n$=0}
\put(95,16){1}
\put(125,22){2}
\put(125,62){3}
\put(95,68){4}
\put(188,82){$\bar\Pt$}
\put(188,40){$\PH$}
\put(188, 2){$\Pt$}
\put( 5, 82){$g$}
\put( 5,  2){$g$}
\end{picture}
}
\hspace*{2em}
{ \unitlength 1pt
\begin{picture}(200,125)(10,-15)
\Gluon(20, 85)( 60, 85){4}{4}
\Gluon(20,  5)( 60,  5){4}{4}
\Gluon(60, 85)(140, 85){4}{8}
\Gluon(60,  5)(140,  5){4}{8}
\Gluon(60, 85)( 60,  5){4}{8}
\Gluon(140, 85)(180, 85){4}{4}
\Gluon(140, 85)(140, 45){4}{4}
\Vertex( 60, 85){3}
\Vertex( 60,  5){3}
\Vertex(140,  5){3}
\Vertex(140, 45){3}
\Vertex(140, 85){3}
\ArrowLine(180,45)(140,45)
\ArrowLine(140,45)(140, 5)
\ArrowLine(140, 5)(180, 5)
\LongArrow(47,55)(47,35)
\LongArrow(149,55)(149,75)
\LongArrow(149,15)(149,35)
\LongArrow(110,97)(90,97)
\LongArrow(90,-5)(110,-5)
\put(36,43){$q$}
\put( 85,-17){$q+p_1$}
\put(155,22){$q+p_1-p_3$}
\put(155,62){$q+p_5-p_2$}
\put( 85,105){$q-p_2$}
\put(69,43){$n$=0}
\put(95,16){1}
\put(125,22){2}
\put(125,62){3}
\put(95,68){4}
\put(188,82){$g$}
\put(188,40){$\bar\Pt$}
\put(188, 2){$\Pt$}
\put( 5, 82){$g$}
\put( 5,  2){$g$}
\end{picture}
}
}
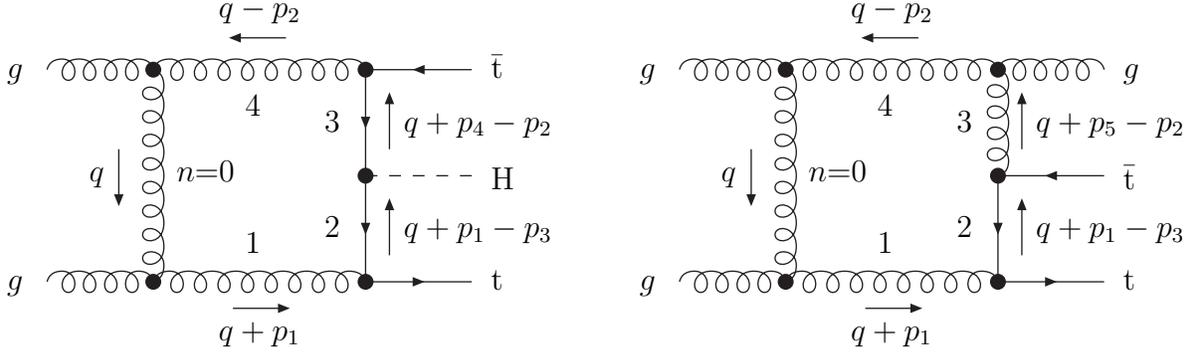
\caption{Examples for pentagon diagrams contributing to the one-loop
corrections to the processes $gg\to\Pt\bar\Pt\PH$ and $gg\to\Pt\bar\Pt g$}
\label{fig:ggtthdiag}
\efi
In order to make use of the direct reduction 
\cite{Me65,Denner:2002ii} of 5-point to
4-point integrals in four space-time dimensions, the dimensionally
regulated integrals have been translated into a mass regularization
by using the trick described in the previous section. However,
the construction of the singular parts of the 5-point in terms
of 3-point integrals has been done integral by integral.
For the 5-point integral on the l.h.s.\ of \reffi{fig:ggtthdiag},
which is a tensor integral of rank 4, we can easily verify that
the explicit formula \refeq{eq:Npointsing} yields the same result
as quoted in \citere{Beenakker:2002nc}.
Assigning the momenta according to
\beq
g(p_1)+g(p_2) \to \Pt(p_3)+\bar \Pt(p_4)+H(p_5)
\eeq
and the defining
\beq
s = (p_1+p_2)^2, \quad s_{ij} = (p_i+p_j)^2, \quad 
t_{kj} = (p_k-p_j)^2, \qquad k=1,2, \quad i,j=3,4,5,
\label{eq:kindef}
\eeq
the auxiliary quantities $\hat d_{mn}$ read
\beq
\left(\hat d_{mn}\right) \;=\;
\pmatrix{
0            & 0            & t_{13}      & t_{24}      & 0 \cr
0            & 0            & \Mt^2       & s_{35}      & s \cr
t_{13}-\Mt^2 & 0            & -\Mt^2      & \MH^2-\Mt^2 & s_{45}-\Mt^2 \cr
t_{24}-\Mt^2 & s_{35}-\Mt^2 & \MH^2-\Mt^2 & -\Mt^2      & 0 \cr
0            & s            & s_{45}      & \Mt^2       & 0},
\eeq
where an obvious matrix notation is used.
Inserting this into Eq.~\refeq{eq:Npointsing} and identifying the singular
3-point integrals yields
\beqar
\lefteqn{
E_{\dots}(0,p_1,p_1-p_3,p_4-p_2,-p_2,0,0,\Mt,\Mt,0)\Big|_{\sing} } 
\qquad
\nn\\*
&=&
\frac{1}{s(s_{35}-\Mt^2)}\,
C_{\dots}(0,p_1,p_1-p_3,0,0,\Mt)
\nn\\ && {}
- \frac{(t_{24}-s_{35})^2}{s(s_{35}-\Mt^2)(t_{13}-\Mt^2)(t_{24}-\Mt^2)}\,
C_{\dots}(0,p_1,p_4-p_2,0,0,\Mt)
\nn\\ && {}
+ \frac{1}{s(s_{45}-\Mt^2)}\,
C_{\dots}(0,p_4-p_2,-p_2,0,\Mt,0)
\nn\\ && {}
+ \frac{1}{(t_{13}-\Mt^2)(t_{24}-\Mt^2)}\,
C_{\dots}(0,p_1,-p_2,0,0,0)
\nn\\ && {}
- \frac{(t_{13}-s_{45})^2}{s(s_{45}-\Mt^2)(t_{13}-\Mt^2)(t_{24}-\Mt^2)}\,
C_{\dots}(0,p_1-p_3,-p_2,0,\Mt,0),
\eeqar
in agreement with Eq.(2.37) of \citere{Beenakker:2002nc}.

\paragraph{{A 5-point integral for the process $gg\to\Pt\bar\Pt g$}}

As another example, we consider the 5-point integral corresponding to
the diagram on the r.h.s.\ of \reffi{fig:ggtthdiag} which contributes to
the (yet unknown) NLO QCD correction to $gg\to\Pt\bar\Pt g$. 
Analogously to the previous case, we assign the momenta according to
\beq
g(p_1)+g(p_2) \to \Pt(p_3)+\bar \Pt(p_4)+g(p_5)
\eeq
and keep the definitions \refeq{eq:kindef}.
The auxiliary quantities $\hat d_{mn}$ read
\beq
\left(\hat d_{mn}\right) \;=\;
\pmatrix{
0            & 0      & t_{13} & t_{25} & 0 \cr
0            & 0      & \Mt^2  & s_{34} & s \cr
t_{13}-\Mt^2 & 0      & -\Mt^2 & 0      & s_{45}-\Mt^2 \cr
t_{25}       & s_{34} & \Mt^2  & 0      & 0 \cr
0            & s      & s_{45} & 0      & 0}.
\eeq
Seven soft or collinear singular 3-point subdiagrams can be 
identified, and Eq.~\refeq{eq:Npointsing} yields
\beqar
\lefteqn{
E_{\dots}(0,p_1,p_1-p_3,p_5-p_2,-p_2,0,0,\Mt,0,0)\Big|_{\sing} } 
\qquad
\nn\\*
&=&
\frac{1}{s s_{34}}\,
C_{\dots}(0,p_1,p_1-p_3,0,0,\Mt)
\nn\\ && {}
- \frac{(t_{25}-s_{34})^2}{s s_{34}t_{25}(t_{13}-\Mt^2)}\,
C_{\dots}(0,p_1,p_5-p_2,0,0,0)
\nn\\ && {}
+\frac{1}{t_{25}s_{34}}\,
C_{\dots}(p_1-p_3,p_5-p_2,-p_2,\Mt,0,0)
\nn\\ && {}
+\frac{1}{s(s_{45}-\Mt^2)}\,
C_{\dots}(0,p_5-p_2,-p_2,0,0,0)
\nn\\ && {}
-\frac{(s-s_{34})^2}{s s_{34} t_{25}(s_{45}-\Mt^2)}\,
C_{\dots}(p_1,p_5-p_2,-p_2,0,0,0)
\nn\\ && {}
+\frac{1}{t_{25}(t_{13}-\Mt^2)}\,
C_{\dots}(0,p_1,-p_2,0,0,0)
\nn\\ && {}
- \frac{(t_{13}-s_{45})^2}{s t_{25}(s_{45}-\Mt^2)(t_{13}-\Mt^2)}\,
C_{\dots}(0,p_1-p_3,-p_2,0,\Mt,0).
\eeqar
This result can again be used to relate the dimensionally regulated
integral into any 4-dimensional regularization which then can be
reduced to 4-point integrals using the method of 
\citeres{Me65,Denner:2002ii}.

\paragraph{{A 5-point integral for the process 
$\Pep\Pem\to\nu_\Pe\bar\nu_\Pe\PH$}}

Now we consider the diagram on the l.h.s.\ of \reffi{fig:eexxhdiag},
which contributes to the ${\cal O}(\alpha)$ corrections of the
process
\beq
\Pem(p_1)+\Pep(p_2) \to \nu_\Pe(p_3)+\bar\nu_\Pe(p_4)+H(p_5).
\eeq
\bfi
\centerline{
{ \unitlength 1pt
\begin{picture}(200,125)(10,-15)
\ArrowLine(60,85)( 20,85)
\ArrowLine(20, 5)( 60, 5)
\ArrowLine(140, 85)(60, 85)
\ArrowLine(60,  5)(140,  5)
\ArrowLine(180, 85)(140, 85)
\ArrowLine(140,  5)(180,  5)
\Photon(60, 85)( 60, 5){3}{8}
\Photon(140,85)(140, 5){3}{8}
\Vertex( 60, 85){3}
\Vertex( 60,  5){3}
\Vertex(140,  5){3}
\Vertex(140, 45){3}
\Vertex(140, 85){3}
\DashLine(140,45)(180,45){5}
\LongArrow(47,55)(47,35)
\LongArrow(149,55)(149,75)
\LongArrow(149,15)(149,35)
\LongArrow(110,97)(90,97)
\LongArrow(90,-5)(110,-5)
\put(36,43){$q$}
\put( 85,-17){$q+p_1$}
\put(155,22){$q+p_1-p_3$}
\put(155,62){$q+p_4-p_2$}
\put( 85,105){$q-p_2$}
\put(69,43){$\gamma(0)$}
\put(85,16){$\Pe(1)$}
\put(110,22){$\PW(2)$}
\put(110,62){$\PW(3)$}
\put(85,68){$\Pe(4)$}
\put(188,82){$\bar\nu_\Pe$}
\put(188,40){$\PH$}
\put(188, 2){$\nu_\Pe$}
\put( 5, 82){$\Pep$}
\put( 5,  2){$\Pem$}
\end{picture}
}
\hspace*{2em}
{ \unitlength 1pt
\begin{picture}(200,125)(10,-15)
\ArrowLine(60,85)( 20,85)
\ArrowLine(20, 5)( 60, 5)
\ArrowLine( 60, 5)(60,85)
\Photon(60,85)(140,85){3}{8}
\Photon(60, 5)(140, 5){3}{8}
\Vertex( 60, 85){3}
\Vertex( 60,  5){3}
\Vertex(140,  5){3}
\Vertex(140, 45){3}
\Vertex(140, 85){3}
\DashLine(140,45)(180,45){5}
\ArrowLine(180, 85)(140, 85)
\ArrowLine(140, 85)(140, 45)
\ArrowLine(140, 45)(140,  5)
\ArrowLine(140,  5)(180,  5)
\LongArrow(47,55)(47,35)
\LongArrow(149,55)(149,75)
\LongArrow(149,15)(149,35)
\LongArrow(110,97)(90,97)
\LongArrow(90,-5)(110,-5)
\put(36,43){$q$}
\put( 85,-17){$q+p_1$}
\put(155,22){$q+p_1-p_3$}
\put(155,62){$q+p_4-p_2$}
\put( 85,105){$q-p_2$}
\put(69,43){$\Pe(0)$}
\put(85,16){$\PZ(1)$}
\put(115,22){$\Pt(2)$}
\put(115,62){$\Pt(3)$}
\put(85,68){$\gamma(4)$}
\put(188,82){$\bar\Pt$}
\put(188,40){$\PH$}
\put(188, 2){$\Pt$}
\put( 5, 82){$\Pep$}
\put( 5,  2){$\Pem$}
\end{picture}
}
}
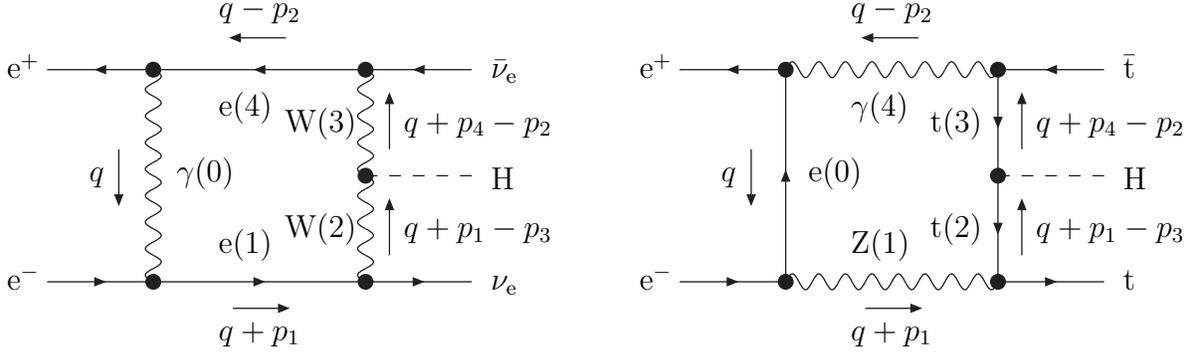
\caption{Examples for pentagon diagrams contributing to the one-loop
corrections to the processes $\Pep\Pem\to\nu_\Pe\bar\nu_\Pe\PH$ and
$\Pep\Pem\to\Pt\bar\Pt\PH$ (The number $n$ of propagator $N_n$ is indicated
in parentheses.)}
\label{fig:eexxhdiag}
\efi
The soft singularity arising from photon exchange is regularized
by the infinitesimally small photon mass $\lambda$. The electron
mass $\Me$ is considered to be much smaller than all other masses and
scalar products, thus leading to mass-singular $\ln\Me$ terms.
Again we make use of definition \refeq{eq:kindef} for the kinematical
variables.
The auxiliary quantities $\hat d_{mn}$ read
\beq
\left(\hat d_{mn}\right) \;=\;
\pmatrix{
0            & 0            & t_{13}      & t_{24}      & 0 \cr
0            & 0            & 0           & s_{35}      & s \cr
t_{13}-\MW^2 & -\MW^2       & -\MW^2      & \MH^2-\MW^2 & s_{45}-\MW^2 \cr
t_{24}-\MW^2 & s_{35}-\MW^2 & \MH^2-\MW^2 & -\MW^2      & -\MW^2 \cr 
0            & s            & s_{45}      & 0           & 0}. 
\eeq
Inserting this into Eq.~\refeq{eq:Npointsing} and identifying the singular
3-point integrals yields
\beqar
\lefteqn{
E_{\dots}(0,p_1,p_1-p_3,p_4-p_2,-p_2,\lambda,\Me,\MW,\MW,\Me)
\Big|_{\sing} } 
\qquad
\nn\\*
&=&
\frac{t_{13}^2}{[(s_{35}t_{13}+\MW^2(t_{24}-t_{13}-s_{35})]s(t_{13}-\MW^2)}\,
C_{\dots}(0,p_1,p_1-p_3,0,\Me,\MW)
\nn\\ && {}
- \frac{(t_{24}-s_{35})^2}{[(s_{35}t_{13}+\MW^2(t_{24}-t_{13}-s_{35})]
s(t_{24}-\MW^2)}\,
C_{\dots}(0,p_1,p_4-p_2,0,\Me,\MW)
\nn\\ && {}
+ \frac{t_{24}^2}{[s_{45} t_{24}+\MW^2(t_{13}-t_{24}-s_{45})]s(t_{24}-\MW^2)}\,
C_{\dots}(0,p_4-p_2,-p_2,0,\MW,\Me)
\nn\\ && {}
+ \frac{1}{(t_{13}-\MW^2)(t_{24}-\MW^2)}\,
C_{\dots}(0,p_1,-p_2,\lambda,\Me,\Me)
\nn\\ && {}
- \frac{(t_{13}-s_{45})^2}{[s_{45} t_{24}+\MW^2(t_{13}-t_{24}-s_{45})]
s(t_{13}-\MW^2)}\,
C_{\dots}(0,p_1-p_3,-p_2,0,\MW,\Me),
\nn\\
\label{eq:eennhpent}
\eeqar
where $\lambda$ is set to zero in all $C_{\dots}$ functions that are not
soft singular.
In \citere{Denner:2003yg}, where the ${\cal O}(\alpha)$ corrections to the
processes $\Pep\Pem\to\nu\bar\nu\PH$ have been worked out,
the 5-point integral has been calculated directly from the corresponding
4-point integrals using the method of \citeres{Me65,Denner:2002ii}.
We have numerically verified the correctness of Eq.~\refeq{eq:eennhpent} by
checking the difference between $E_0$ taken from \citere{Denner:2003yg}
and the r.h.s.\ of Eq.~\refeq{eq:eennhpent} to be independent
of $\ln\lambda$ and $\ln\Me$.

\paragraph{{A 5-point integral for the process $\Pep\Pem\to\Pt\bar\Pt\PH$}}

Finally, we turn to the diagram on the r.h.s.\ of \reffi{fig:eexxhdiag},
which contributes to
\beq
\Pem(p_1)+\Pep(p_2) \to \Pt(p_3)+\bar \Pt(p_4)+H(p_5)
\eeq
at one loop. The masses $\lambda$ and $\Me$ are treated as in the 
previous example, and
definition \refeq{eq:kindef} is again employed.
The auxiliary quantities $\hat d_{mn}$ read
\beq
\left(\hat d_{mn}\right) \;=\;
\pmatrix{
0            & 0            & t_{13}      & t_{24}       & 0 \cr
-\MZ^2       & -\MZ^2       & \Mt^2-\MZ^2 & s_{35}-\MZ^2 & s-\MZ^2 \cr
t_{13}-\Mt^2 & 0            & -\Mt^2      & \MH^2-\Mt^2  & s_{45}-\Mt^2 \cr
t_{24}-\Mt^2 & s_{35}-\Mt^2 & \MH^2-\Mt^2 & -\Mt^2       & 0 \cr 
0            & s            & s_{45}      & \Mt^2        & 0}. 
\eeq
Equation~\refeq{eq:Npointsing} yields
\beqar
\lefteqn{
E_{\dots}(0,p_1,p_1-p_3,p_4-p_2,-p_2,\Me,\MZ,\Mt,\Mt,\lambda)
\Big|_{\sing} } 
\qquad
\nn\\*
&=&
\frac{1}{(s-\MZ^2)(s_{45}-\Mt^2)}\,
C_{\dots}(0,p_4-p_2,-p_2,\Me,\Mt,\lambda)
\nn\\ && {}
+ \frac{s^2}{[s(t_{13}-\Mt^2)+\MZ^2(s_{45}-t_{13})](s-\MZ^2)(t_{24}-\Mt^2)}\,
C_{\dots}(0,p_1,-p_2,\Me,\MZ,0)
\nn\\ && {}
- \frac{(t_{13}-s_{45})^2}{[s(t_{13}-\Mt^2)+\MZ^2(s_{45}-t_{13})]
(s_{45}-\Mt^2)(t_{24}-\Mt^2)}\,
\nn\\[.3em] && {} \quad \times
C_{\dots}(0,p_1-p_3,-p_2,\Me,\Mt,0).
\eeqar
This relation has been numerically checked against the results in the 
calculation
\cite{Denner:2003ri} of the ${\cal O}(\alpha)$ corrections to the
process $\Pep\Pem\to\Pt\bar\Pt\PH$, as described in the previous example.

\section{Summary}
\label{se:summary}

By analyzing the soft and collinear limits in one-loop 
$N$-point integrals, the mass singularities of general tensor
one-loop $N$-point integrals are explicitly expressed in terms of
3-point integrals. The remarkably simple result is illustrated
in some non-trivial examples, such as 5-point integrals
occurring in $2\to3$ particle reactions.

The final formula holds true in
any regularization scheme and, thus, can not only
be used to predict the soft and collinear singularities of integrals,
but also to translate mass-singular one-loop integrals from one
regularization scheme to another. This, in particular, 
shows that the reduction of $N$-point integrals with $N\ge5$ to
4-point integrals in $D$ dimensions proceeds as in four dimensions
without extra contributions.

Since the separation of soft and collinear singularities is carried out
in momentum space, the method can also be used as subtraction formalism
for one-loop integrals, similar to the subtraction approaches used
in the calculation of real radiative corrections.

\appendix
\section*{Appendix}

\section{Proof of an auxiliary identity}
\label{app:partialfrac}

Here we make up for the proof of Eq.~\refeq{eq:Ank}. 
We first proof the auxiliary identity
\beq
\sum_{k=0}^N \;\frac{b_k^{N-1}}
{\displaystyle\prod_{l=0 \atop l\ne k}^N (b_k a_l-b_l a_k)}=0,
\label{eq:auxrel}
\eeq
where $a_k$, $b_k$ ($k=0,1,\dots$) represent any two series of 
variables so that the denominators in the above expressions are non-zero.
We proceed by induction. The case $N=1$ is trivial, since then the sum
consists of two terms with opposite sign. 
Now we assume Eq.~\refeq{eq:auxrel} to be valid and deduce the analogous 
relation for $N\to N+1$, i.e.\ we have to show that the following
expression vanishes,
\beqar
\sum_{k=0}^{N+1} \;\frac{b_k^N}
{\displaystyle\prod_{l=0 \atop l\ne k}^{N+1} (b_k a_l-b_l a_k)}
&=&
\sum_{k=0}^{N-1} \;
\frac{b_k}{(b_k a_N-b_N a_k)(b_k a_{N+1}-b_{N+1} a_k)} \,
\frac{b_k^{N-1}}{\displaystyle\prod_{l=0 \atop l\ne k}^{N-1} (b_k a_l-b_l a_k)}
\nn\\ && {}
+ \frac{b_N^N}{(b_N a_{N+1}-b_{N+1} a_N)
\displaystyle\prod_{l=0}^{N-1} (b_k a_l-b_l a_k)}
\nn\\ && {}
+ \frac{b_{N+1}^N}{(b_{N+1} a_N-b_N a_{N+1})
\displaystyle\prod_{l=0}^{N-1} (b_k a_l-b_l a_k)}.
\eeqar
In the last equation we have only separated the last two terms from the sum 
and extracted some factors for convenience. Now we rewrite a factor
of the first term on the r.h.s.\ as follows,
\beqar
\lefteqn{ \hspace{-2em}
\frac{b_k}{(b_k a_N-b_N a_k)(b_k a_{N+1}-b_{N+1} a_k)} }
\nn\\
&=&
\frac{1}{b_N a_{N+1}-b_{N+1} a_N} \left(
\frac{b_N}{b_k a_N-b_N a_k} - \frac{b_{N+1}}{b_k a_{N+1}-b_{N+1} a_k} 
\right),
\eeqar
and obtain
\beqar
\lefteqn{ \sum_{k=0}^{N+1} \;\frac{b_k^N}
{\displaystyle\prod_{l=0 \atop l\ne k}^{N+1} (b_k a_l-b_l a_k)} }
\nn\\
&=&
\frac{1}{b_N a_{N+1}-b_{N+1} a_N} \Biggl[
\sum_{k=0}^{N-1}
\left(
\frac{b_N}{b_k a_N-b_N a_k} - \frac{b_{N+1}}{b_k a_{N+1}-b_{N+1} a_k}     
\right)
\frac{b_k^{N-1}}{\displaystyle\prod_{l=0 \atop l\ne k}^{N-1} (b_k a_l-b_l a_k)}
\nn\\
&& \qquad {} 
+ \frac{b_N^N}{\displaystyle\prod_{l=0}^{N-1} (b_k a_l-b_l a_k)}
- \frac{b_{N+1}^N}{\displaystyle\prod_{l=0}^{N-1} (b_k a_l-b_l a_k)}
\Biggr]
\nn\\
&=&
\frac{1}{b_N a_{N+1}-b_{N+1} a_N} \Biggl[
b_N \sum_{k=0}^N \,
\frac{b_k^{N-1}}{\displaystyle\prod_{l=0 \atop l\ne k}^N (b_k a_l-b_l a_k)}
- b_{N+1} \sum_{k=0 \atop k\ne N}^{N+1} \,
\frac{b_k^{N-1}}{\displaystyle\prod_{l=0 \atop l\ne k,N}^{N+1}(b_k a_l-b_l a_k)}
\Biggr]
\nn\\
&=& 0.
\eeqar
Each of the sum in the last-but-one line vanishes by assuming the
validity of Eq.~\refeq{eq:auxrel}. For the first sum this is obvious,
for the second it should be realized that the sum follows from 
Eq.~\refeq{eq:auxrel} by the substitutions $a_N\to a_{N+1}$ and
$b_N\to b_{N+1}$.

In order to proof Eq.~\refeq{eq:Ank}, we have to show
\beq
\prod_{k=0 \atop k\ne n,n+1}^{N-1} \frac{1}{- x_n \hat c_{kn} + \hat d_{kn}} 
= \sum_{k=0 \atop k\ne n,n+1}^{N-1} 
\frac{\hat c_{kn}^{N-3}}{\disp\prod_{l=0 \atop l\ne k,n,n+1}^{N-1}
(\hat c_{kn}\hat d_{ln}-\hat c_{ln}\hat d_{kn})} \,
\frac{1}{- x_n \hat c_{kn} + \hat d_{kn}}.
\label{eq:Ankrel}
\eeq
Again we proceed by induction. For $N=3$, there is nothing to show, as
the l.h.s.\ and the r.h.s.\ are obviously equal.
Now we assume the validity of Eq.~\refeq{eq:Ankrel} and proceed from
$N$ to $N+1$ terms in the product on the l.h.s.,
\beqar
\prod_{k=0 \atop k\ne n,n+1}^N \frac{1}{- x_n \hat c_{kn} + \hat d_{kn}}
&=&
\frac{1}{- x_n \hat c_{Nn} + \hat d_{Nn}}
\prod_{k=0 \atop k\ne n,n+1}^{N-1} \frac{1}{- x_n \hat c_{kn} + \hat d_{kn}}
\nn\\
&=&
\sum_{k=0 \atop k\ne n,n+1}^{N-1} 
\frac{\hat c_{kn}^{N-3}}{\disp\prod_{l=0 \atop l\ne k,n,n+1}^{N-1}
(\hat c_{kn}\hat d_{ln}-\hat c_{ln}\hat d_{kn})} \,
\frac{1}{(- x_n \hat c_{kn} + \hat d_{kn})
(- x_n \hat c_{Nn} + \hat d_{Nn})}.
\nn\\
\eeqar
The last equation follows from assuming Eq.~\refeq{eq:Ankrel}.
Note that we have assumed $n\ne N,N-1$ in the first manipulation,
which is no loss of generality, since we can always achieve this by
renumbering the propagators.
Next we use the partial fraction
\beq
\frac{1}{(- x_n \hat c_{kn} + \hat d_{kn})
(- x_n \hat c_{Nn} + \hat d_{Nn})} = 
\frac{1}{\hat c_{kn}\hat d_{Nn}-\hat c_{Nn}\hat d_{kn}} \left[
\frac{\hat c_{kn}}{- x_n \hat c_{kn} + \hat d_{kn}}
-\frac{\hat c_{Nn}}{- x_n \hat c_{Nn} + \hat d_{Nn}} \right]
\eeq
to obtain
\beqar
\prod_{k=0 \atop k\ne n,n+1}^N \frac{1}{- x_n \hat c_{kn} + \hat d_{kn}}
&=&
\sum_{k=0 \atop k\ne n,n+1}^{N-1} 
\frac{\hat c_{kn}^{N-2}}{\disp\prod_{l=0 \atop l\ne k,n,n+1}^N
(\hat c_{kn}\hat d_{ln}-\hat c_{ln}\hat d_{kn})} \,
\frac{1}{- x_n \hat c_{kn} + \hat d_{kn}}
\nn\\
&& {}
- \frac{\hat c_{Nn}}{- x_n \hat c_{Nn} + \hat d_{Nn}}
\sum_{k=0 \atop k\ne n,n+1}^{N-1} 
\frac{\hat c_{kn}^{N-3}}{\disp\prod_{l=0 \atop l\ne k,n,n+1}^N
(\hat c_{kn}\hat d_{ln}-\hat c_{ln}\hat d_{kn})} 
\nn\\
&=&
\sum_{k=0 \atop k\ne n,n+1}^N 
\frac{\hat c_{kn}^{N-2}}{\disp\prod_{l=0 \atop l\ne k,n,n+1}^N
(\hat c_{kn}\hat d_{ln}-\hat c_{ln}\hat d_{kn})} \,
\frac{1}{- x_n \hat c_{kn} + \hat d_{kn}}
\nn\\
&& {}
- \frac{\hat c_{Nn}}{- x_n \hat c_{Nn} + \hat d_{Nn}}
\sum_{k=0 \atop k\ne n,n+1}^N
\frac{\hat c_{kn}^{N-3}}{\disp\prod_{l=0 \atop l\ne k,n,n+1}^N
(\hat c_{kn}\hat d_{ln}-\hat c_{ln}\hat d_{kn})}.
\eeqar
The last equality follows from the fact that the terms with $k=N$ in the two 
sums cancel each other. Note that the last sum is identically zero
owing to auxiliary relation \refeq{eq:auxrel} proven above.
This completes the proof of Eqs.~\refeq{eq:Ankrel} and \refeq{eq:Ank}.

\section{Mass-singular scalar 3-point functions}
\label{app:C0s}

\newcommand{\CO}[6]{\raisebox{-3.0em}{
{ \unitlength 1pt
\begin{picture}(130,70)(0,10)
\Line(20,45)( 50,45)
\Line(50,45)( 80,65)
\Line(80,65)(110,75)
\Line(50,45)( 80,25)
\Line(80,25)(110,15)
\Line(80,25)( 80,65)
\Vertex( 50, 45){3}
\Vertex( 80, 65){3}
\Vertex( 80, 25){3}
\Text( 15, 45)[r]{$#1$}
\Text(115, 20)[l]{$#2$}
\Text(115, 70)[l]{$#3$}
\Text( 65, 65)[r]{$#4$}
\Text( 65, 25)[r]{$#5$}
\Text( 85, 45)[l]{$#6$}
\end{picture}
} }}
In this appendix we give a list of mass-singular scalar 3-point integrals
that frequently appear in applications. Whenever the mass parameter
$\lambda$ appears, it is understood as infinitesimal. 
If not all singularities are cured by mass parameters, dimensional
regularization is applied.
Thus, in the following all formulas are valid up to order 
${\cal O}(\lambda)$ or ${\cal O}(\eps)$.
For convenience, a graphical notation is used,
\beq
\CO{(p_1-p_0)^2}{(p_2-p_1)^2}{(p_0-p_2)^2}{m_0}{m_1}{m_2} \qquad
\equiv \; C_0(p_0,p_1,p_2,m_0,m_1,m_2),
\eeq
and overlined variables are understood to receive an infinitesimally
small imaginary part, $\bar s=s+\ri0$, etc.
The function $\Li(x)=-\int^x_0 \rd t \ln(1-t)/t$ denotes the usual
dilogarithm.

Collinear singularities show up in the following cases:
\beqar
\hspace{-1em}
\CO{s_1}{s_2}{\lambda^2}{\lambda}{m}{0} \quad &=&
\frac{1}{s_1-s_2} \Biggl\{
 \ln\left(\frac{m^2-\bar s_1}{\lambda^2}\right)
 \ln\left(\frac{m^2-\bar s_1}{m^2}\right)
\nn\\[-1.5em]
&& {}
-\ln\left(\frac{m^2-\bar s_2}{\lambda^2}\right)
 \ln\left(\frac{m^2-\bar s_2}{m^2}\right)
-2\Li\left(\frac{\bar s_1-\bar s_2}{m^2-\bar s_2}\right) 
\nn\\[.5em]
&& {}
+\Li\left(\frac{\bar s_1}{m^2}\right)
-\Li\left(\frac{\bar s_2}{m^2}\right) 
\Biggr\},
\\
\hspace{-1em}
\CO{s_1}{s_2}{0}{\lambda}{m}{\lambda} \quad &=&
\frac{1}{s_1-s_2} \Biggl\{
\ln^2\left(\frac{m^2-\bar s_1}{\lambda m}\right)
-\ln^2\left(\frac{m^2-\bar s_2}{\lambda m}\right)
\nn\\[-1.5em]
&& {}
+\Li\left(\frac{\bar s_1}{m^2}\right)
-\Li\left(\frac{\bar s_2}{m^2}\right) \Biggr\},
\\
\hspace{-1em}
\CO{s_1}{s_2}{0}{0}{m}{0} \quad &=&
\frac{1}{s_1-s_2} \Biggl\{
\frac{\Gamma(1+\eps)}{\eps}\left(\frac{4\pi\mu^2}{m^2}\right)^\eps
\ln\left(\frac{m^2-\bar s_2}{m^2-\bar s_1}\right)
\nn\\[-1.5em]
&& {}
+\ln^2\left(\frac{m^2-\bar s_1}{m^2}\right)
-\ln^2\left(\frac{m^2-\bar s_2}{m^2}\right)
\nn\\[.5em]
&& {}
+\Li\left(\frac{\bar s_1}{m^2}\right)
-\Li\left(\frac{\bar s_2}{m^2}\right) + {\cal O}(\eps) \Biggr\},
\eeqar

There is only one situation for a pure soft singularity
(given in two regularization schemes):
\beqar
\hspace{-1em}
\CO{s}{m_1^2}{m_2^2}{m_2}{m_1}{\lambda} \quad &=&
\frac{x_s}{m_1 m_2 (1-x_s^2)} \biggl\{
-\ln\left(\frac{\lambda^2}{m_1 m_2}\right)\ln(x_s)
\nn\\[-1.5em]
&& {} 
-\frac{1}{2}\ln^2(x_s)
+2\ln(x_s)\ln(1-x_s^2)
+\frac{1}{2}\ln^2\left(\frac{m_1}{m_2}\right)
-\frac{\pi^2}{6} 
\nn\\[.5em]
&& {}
+\Li(x_s^2)
+\Li\left(1-x_s\frac{m_1}{m_2}\right)
+\Li\left(1-x_s\frac{m_2}{m_1}\right) \biggr\},
\hspace{2em}
\label{eq:C0soft}
\\
\hspace{-1em}
\CO{s}{m_1^2}{m_2^2}{m_2}{m_1}{0} \quad &=&
\frac{x_s}{m_1 m_2 (1-x_s^2)} \biggl\{
-\frac{\Gamma(1+\eps)}{\eps}\left(\frac{4\pi\mu^2}{m_1 m_2}\right)^\eps\ln(x_s)
-\frac{1}{2}\ln^2(x_s)
\nn\\[-1.5em]
&& {} 
+2\ln(x_s)\ln(1-x_s^2)
+\frac{1}{2}\ln^2\left(\frac{m_1}{m_2}\right)
-\frac{\pi^2}{6} 
+\Li(x_s^2)
\nn\\[.5em]
&& {}
+\Li\left(1-x_s\frac{m_1}{m_2}\right)
+\Li\left(1-x_s\frac{m_2}{m_1}\right) + {\cal O}(\eps) \biggr\},
\hspace{2em}
\eeqar
where Eq.~\refeq{eq:C0soft} is taken over from \citere{Beenakker:1990jr}
with
\beq
x_s = \frac{\sqrt{1-4m_1 m_2/\Big[\bar s-(m_1-m_2)^2\Big]}-1}
{\sqrt{1-4m_1 m_2/\Big[\bar s-(m_1-m_2)^2\Big]}+1}.
\eeq

Finally, collinear and soft singularities can overlap:
\beqar
\hspace{-1em}
\CO{s}{m_1^2\ll |s|,m_2^2}{m_2^2}{m_2}{m_1}{\lambda} \quad &=&
\frac{1}{s-m_2^2} \Biggl\{
\ln\left(\frac{m_1(m_2^2-\bar s)}{\lambda^2 m_2}\right)
\ln\left(\frac{m_2^2-\bar s}{m_1 m_2}\right)
+\Li\left(\frac{\bar s}{m_2^2}\right) \Biggr\},
\nn\\[-1.5em]
\\[1em]
\hspace{-1em}
\CO{s}{m_1^2\ll |s|}{m_2^2\ll |s|}{m_2}{m_1}{\lambda} \quad &=&
\frac{1}{s} \Biggl\{
\ln\left(\frac{\lambda^2}{-\bar s}\right)\ln\left(\frac{m_1 m_2}{-\bar s}\right)
\nn\\*[-1.5em]
&& {} 
\qquad
-\frac{1}{4}\ln^2\left(\frac{m_1^2}{-\bar s}\right)
-\frac{1}{4}\ln^2\left(\frac{m_2^2}{-\bar s}\right)
-\frac{\pi^2}{6} \Biggr\},
\hspace{2em}
\\
\hspace{-1em}
\CO{s}{m^2}{\lambda^2}{0}{m}{\lambda} \quad &=&
\frac{1}{s-m^2} \Biggl\{
 \ln^2\left(\frac{m^2-\bar s}{\lambda m}\right)
+\frac{5\pi^2}{12}
+\Li\left(\frac{\bar s}{m^2}\right)
\Biggr\},
\\
\hspace{-1em}
\CO{s}{m^2}{0}{\lambda}{m}{\lambda} \quad &=&
\frac{1}{s-m^2} \Biggl\{
 \ln^2\left(\frac{m^2-\bar s}{\lambda m}\right)
+\frac{\pi^2}{12}
+\Li\left(\frac{\bar s}{m^2}\right)
\Biggr\},
\\
\hspace{-1em}
\CO{s}{m^2}{0}{0}{m}{0} \quad &=&
\frac{1}{s-m^2} \Biggl\{
 \frac{\Gamma(1+\eps)}{\eps^2}\left(\frac{4\pi\mu^2}{m^2-\bar s}\right)^\eps
-\frac{\Gamma(1+\eps)}{2\eps^2}\left(\frac{4\pi\mu^2}{m^2}\right)^\eps
\nn\\[-1.5em]
&& {}
-\Li\left(\frac{\bar s}{\bar s-m^2}\right) + {\cal O}(\eps)
\Biggr\},
\\
\hspace{-1em}
\CO{s}{\lambda^2}{\lambda^2}{0}{0}{\lambda} \quad &=&
\frac{1}{s} \Biggl\{
\frac{1}{2}\ln^2\left(\frac{\lambda^2}{-\bar s}\right)+\frac{2\pi^2}{3}
\Biggr\},
\\
\hspace{-1em}
\CO{s}{0}{\lambda^2}{0}{\lambda}{\lambda} \quad &=&
\frac{1}{s} \Biggl\{
\frac{1}{2}\ln^2\left(\frac{\lambda^2}{-\bar s}\right)+\frac{\pi^2}{3}
\Biggr\},
\\
\hspace{-1em}
\CO{s}{0}{0}{\lambda}{\lambda}{\lambda} \quad &=&
\frac{1}{2s} \ln^2\left(\frac{\lambda^2}{-\bar s}\right),
\\
\hspace{-1em}
\CO{s}{0}{0}{0}{0}{0} \quad &=&
\frac{1}{s} \Biggl\{
\frac{\Gamma(1+\eps)}{\eps^2}\left(\frac{4\pi\mu^2}{-\bar s}\right)^\eps
-\frac{\pi^2}{6} + {\cal O}(\eps) \Biggr\}.
\eeqar

\section*{Acknowledgement}

The author is grateful to W.~Beenakker for his collaboration when
working out the pentagon integrals of \citere{Beenakker:2002nc};
the methods developed there served as the starting point of this work.
Moreover, he and A.~Denner are acknowledged for valuable discussions
and for carefully reading the manuscript.

\end{document}